\documentclass[           
 ,draft            
  ]
  {aipproc}

\layoutstyle{6x9}

\usepackage{graphicx}
\usepackage{epsfig}
\begin{document}

\newcommand{\be}[1]{\begin{equation}\label{#1}}
 \newcommand{\ee}{\end{equation}}

\title{Evolution of relativistic jets from XTE J1550-564
and the environment of microquasars}


\author{Shuang Nan Zhang}{
  address={Physics Department and Center for Astrophysics,
Tsinghua University, Beijing, 100084, China\footnote{email:
zhangsn@tsinghua.edu.cn, jingfang.hao@hotmail.com}} }

\author{JingFang Hao}{}

\begin{abstract}
Two relativistic X-ray jets have been detected with the \textit{Chandra} X-ray
observatory in the black hole X-ray transient XTE J1550-564. We report a full analysis
of the evolution of the two jets with a gamma-ray burst external shock model. A
plausible scenario suggests a cavity outside the central source and the jets first
travelled with constant velocity and then are slowed down by the interactions between
the jets and the interstellar medium (ISM). The best fitted radius of the cavity is
$\sim$0.36 pc on the eastern side and $\sim$0.46 pc on the western side, and the
densities also show asymmetry, of $\sim$0.015 cm$^{-3}$ on the east to $\sim$0.21
cm$^{-3}$ on the west. Large scale low density region is also found in another
microquasar system, H 1743-322. These results are consistent with previous suggestions
that the environment of microquasars should be rather vacuous, compared to the normal
Galactic environment. A generic scenario for microquasar jets is proposed, classifying
the observed jets into three main categories, with different jet morphologies (and
sizes) corresponding to different scales of vacuous environments surrounding them.
\end{abstract}

\maketitle


\section{Introduction}

Microquasars are well known miniatures of quasars, with a central
black hole (BH), an accretion disk and two relativistic jets very
similar to those found in the centers of active galaxies, only on
much smaller scales (Mirabel $\&$ Radr\'{\i}guez 1999).

Since discovered in 1992, radio jets have been observed in several BH binary systems
and some of them showed apparent superluminal features. In the two well known
microqusars, GRS 1915+105 (Mirabel $\&$ Radr\'{\i}guez 1999) and GRO J1655-40 (Tingay
et al.1995; Hjellming $\&$ Rupen 1995), relativistic jets with actual velocities
greater than 0.9$c$ were observed. In some other systems, small-size ``compact jets",
e.g. Cyg X-1 (Stirling et al. 2001), and large scale diffuse emission, e.g. SS433
(Dubner et al. 1998), were also detected.

XTE J1550-564 was discovered with RXTE in 1998 during its strong X-ray outburst on
September 7 (Smith 1998). It is believed to be an X-ray binary system at a distance of
$\sim$5.3 kpc, containing a black hole of 10.5$\pm$1.0 solar masses and a low mass
companion star (Orosz et al. 2002). Soon after the discovery of the source, a jet
ejection with an apparent velocity greater than 2$c$ was reported (Hannikainen et al.
2001). In the period between 1998 and 2002, several other outbursts occurred but no
similar radio and X-ray flares were detected again in these outbursts (Tomsick et al.
2003).

With the help of the \textit{Chandra} satellite, Corbel et al (2002)
found two large scale X-ray jets lying to the east and the west of
the central source, which were also in good alignment with the
central source. The eastern jet has been detected first in 2000 at a
projected distance of $\sim$21 arcsec from the central black hole.
Two years later, it could only be seen marginally in the X-ray
image, while a western counterpart became visible at $\sim$22 arcsec
on the other side. The corresponding radio maps are consistent with
the X-ray observations (Corbel et al. 2002).

There are altogether eight 2-dimentional imaging observations of XTE J1550-564 in
\textit{Chandra} archive during June 2000 and October 2003 (henceforth observations
1$\sim$8). Here we report a full analysis of these X-ray data, together with the
kinematic and spectral evolution fittings for all these observations.

\section{OBSERVATIONS of XTE J1550-564}

The basic information of observations 1$\sim$8 is listed in Table 1, including the
observation ID, date, and the angular separation between the eastern and western jets
and the central source. The positions are obtained by the \textit{Chandra} Interactive
Analysis of Observations (CIAO) routine $wavdetect$ (Freeman et al. 2002). In
observations 5 and 6, no X-ray source is detected by $wavdetect$ at the position of the
eastern jet. However, from the smoothed images (Fig.1), a weak source could be
recognized in observation 6. We thus select the center of the strongest emission region
as the position of the jet in that observation. We calculate the source centroid for
the central source and the X-ray jet respectively and for all the five observations,
the calculated position changed by less than 0.5$^{\prime}$$^{\prime}$. Therefore, an
upper limit of 0.5$^{\prime}$$^{\prime}$ is set for the error of the jet distance.

\begin{figure}
\begin{minipage}[b]{.5\textwidth}
    \centering
    \includegraphics[scale=0.3]{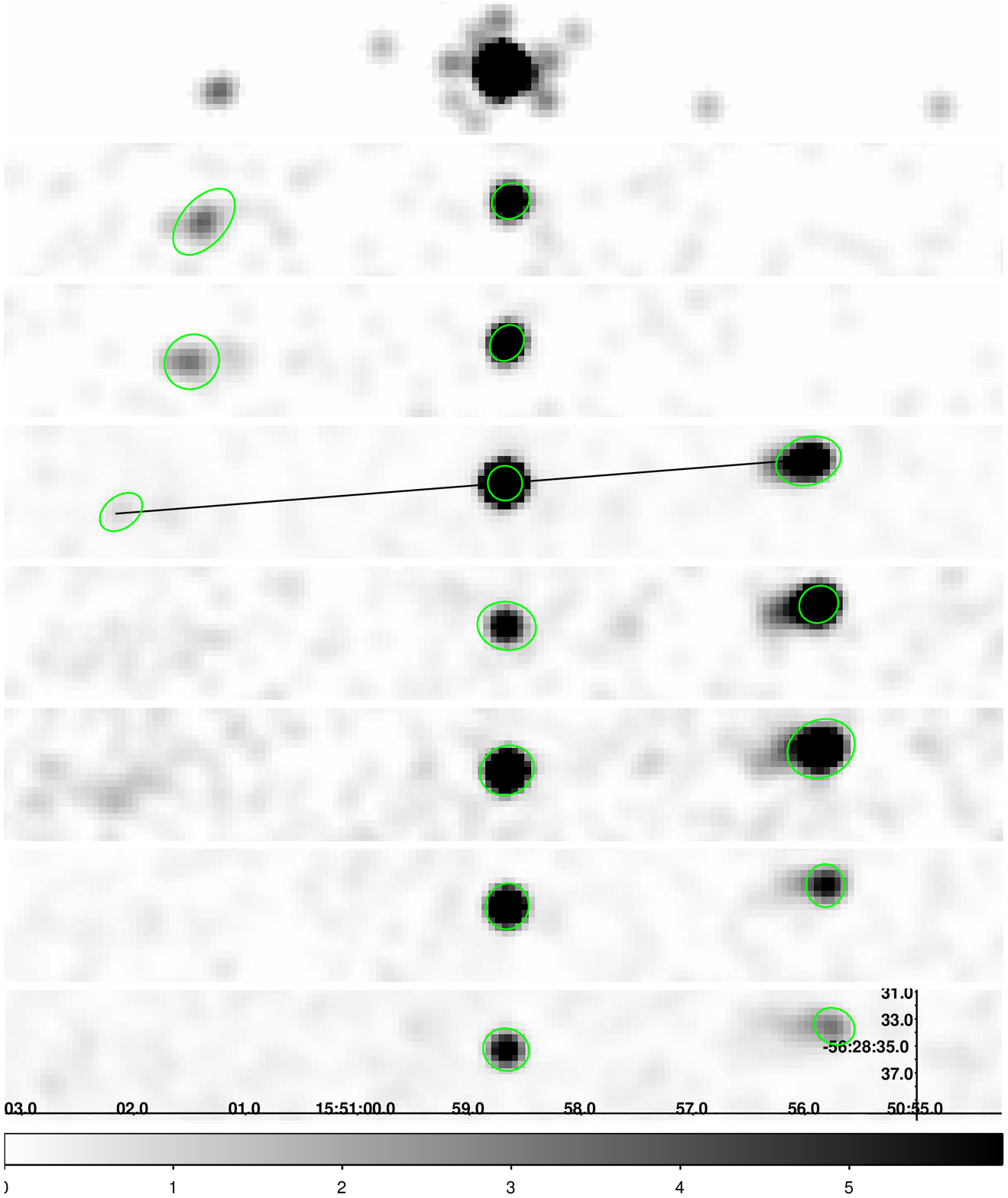}
  \end{minipage}%
  \begin{minipage}[b]{.5\textwidth}
{\bf FIGURE 1.} The smoothed \textit{Chandra} X-ray images of the eight observations of
XTE J1550-564 and the two jets together. The green elliptical regions are source
emission regions by $wavdect$. Observation 4 shows the good alignment of the two jets
and the central source.
  \end{minipage}

\end{figure}

From Table 1 and Fig.1, we could see clearly that an X-ray emission source is detected
to the east of the central source in the first four observations and another source is
detected to the west in the last five observations. Calculations also show that these
two sources, when presented in a single combined image, are in good alignment with the
central compact object with an inclination angle of
85.9\textordmasculine$\pm$0.3\textordmasculine. By calculating the average proper
motion, an approximate estimate of deceleration could be seen for both jets.

\section{ENERGY SPECTRUM and FLUX}

Since the emission from the eastern jet has been studied fully (Corbel et al. 2002;
Tomsick et al. 2003), we mainly focus our spectral analysis on the western jet. The
X-ray spectrum in 0.3-8 keV energy band is extracted for each observation of the
western jet. We use a circular source region with a radius of 4$^{\prime}$$^{\prime}$,
an annular background region with an inner radius of 5$^{\prime}$$^{\prime}$ and an
outer radius of 15$^{\prime}$$^{\prime}$, for each observation. Instrument response
matrices (rmf) and weighted auxiliary response files (warf) are created using CIAO
programs mkacisrmf and mkwarf, and then added to the spectra. We re-bin the spectra
with 10 counts per bin and fit them in \textit{Xspec}.

The results of spectra fitting with an absorbed power-law model are
also shown in table 1. We use the Cash statistic since it is a
better method when counts are low. The absorption column density is
fixed to the Galactic value in the direction of XTE J1550-564
obtained by the radio observations ($N_{H}=9\times10^{21}$cm$^{-2}$
) (Dickey \& Lockman 1990). Our results are quite consistent with
previous works by Karret et al.(2003). The calculated absorbed
energy flux in 0.3-8 keV band is comparable to the value of the
eastern jet. The observed flux decayed rather quickly, from
$\sim1.9\times10^{-13}$erg cm$^{-2}$ s$^{-1}$ in March 2002 to only
one sixth of this value in October 2003 (see section 4.2).

\begin{table}
  \centering
  \caption{XTE J1550-564 \textit{Chandra} Observations}\label{table1}
\begin{tabular}{ccccccc}
\hline
 &  &  &\multicolumn{2}{c}{Angular Separations (arcsec)} &\multicolumn{2}{c}{Powerlaw Fitting for the western jet} \\
\cline{4-5} \cline{6-7}\\
Num & ID   & Date    & Eastern Jet & Western jet    &Photon Index  & Flux (ergs cm$^{-2}$ s$^{-1}$) \\
\hline
1 &679  & 2000 06 09  &21.5$\pm$0.5   &  &&  \\
2 &1845 & 2000 08 21 &22.8$\pm$0.5   &    &  &\\
3 &1846 & 2000 09 11 &23.4$\pm$0.5   &    &  &  \\
4 &3448 & 2002 03 11 &28.6$\pm$0.5   &22.6$\pm$0.5   &1.75$\pm$0.11    &$(1.9\pm0.4)\times10^{-13}$\\
5 &3672 & 2002 06 19 &     &23.2$\pm$0.5   &1.71$\pm$0.15    &$(1.6\pm0.3)\times10^{-13}$ \\
6 &3807 & 2002 09 24 &29.2$\pm$0.5  &23.4$\pm$0.5  &1.94$\pm$0.17    &$(8.6\pm1.5)\times10^{-14}$\\
7 &4368 & 2003 01 28 &     &23.7$\pm$0.5   &1.81$\pm$0.22    &$(5.5\pm1.0)\times10^{-14}$ \\
8 &5190 & 2003 10 23 &    &24.5$\pm$0.5   &1.97$\pm$0.20    &$(3.1\pm0.6)\times10^{-14}$\\
\hline
\end{tabular}
\end{table}

\section{JET MODEL}

\subsection{Kinematic Model}

In the external shock model for afterglows of GRBs, the kinematic
and radiation evolution could be understood as the interaction
between the outburst ejecta and the surrounding ISM. Microquasar jet
systems are also expected to encounter such interactions. In this
section, we describe our attempts after Wang et al. (2003) in
constructing the kinetic and radiation model based on these
theories.

We adopt the model of a collimated conical beam with a half opening
angle $\theta_{j}$ expanding into the ambient medium with the number
density $n$. The initial kinetic energy and Lorentz factor of the
outflow material are $E_{0}$ and $\Gamma_{0}$ , respectively. Shocks
should arise as the outflow moves on and heat the ISM, and its
kinetic energy will turn into the internal energy of the medium
gradually. Neglect the radiation loss, the energy conservation
function writes (Huang, Dai, \& Lu 1999):
\begin{equation}\label{a}
(\Gamma-1)M_{0}c^{2}+\sigma(\Gamma_{\textrm{\tiny
sh}}^{2}-1)m_{\textrm{\tiny SW}}c^{2}=E_{0}
\end{equation}

The first term on the left of the equation represents the kinematic
energy of the ejecta, where $\Gamma$ is the Lorentz factor and
$M_{0}$ is the mass of the original ejecta. The second term
represent the internal energy of the swept-up ISM, where
$\Gamma_{\textrm{\tiny sh}}$ and $m_{\textrm{\tiny SW}}$ are the
corresponding Lorentz Factor and mass of the shocked ISM
respectively, and $ m_{\textrm{\tiny
SW}}=(4/3){\pi}R^{3}m_{\textrm{\tiny p}}n(\theta_{j}^{2}/4)$.

Coefficient $\sigma$ differs from 6/17 to 0.73 for ultrarelativistic
and nonrelativistic jets. We adopt the approximation of
$\sigma\sim$0.7 after Wang et al.(2003). Equation (1) and the
relativistic kinematic equations
\begin{equation}
(\frac{dR}{dt})_{\textrm{a}}=\frac{\beta(\Gamma)c}{1-\beta(\Gamma)\cos\theta};
(\frac{dR}{dt})_{\textrm{r}}=\frac{\beta(\Gamma)c}{1+\beta(\Gamma)\cos\theta}
\end{equation}
can be solved and give the relation between the projected angular separation $\mu$ and
time $t$. In equation (2), the subscript a and r represent the approaching and receding
jets in a pair of relativistic jets respectively. $R$ is the distance between the jet
and the source, which can be transformed into the proper motion separation by
$\mu=R\sin\theta/5.3$ kpc, and $\theta$ is the jet inclination angle to the line of
sight. We can get the $\mu-t$ curve numerically with the above equations. To be
consistent with the work done to the eastern jet, we choose the same initial conditions
that $\Gamma_{0}=3$, $E_{0}=3.6\times10^{44}$ erg, and
$\theta_{j}=1.\textordmasculine5$. Then the parameters needed to be fit are $n$ and
$\theta$.

In the case of the eastern jet, the number density of the ISM was assumed as a constant
in the whole region outside the central source. This assumption does not work well in
the case of its western counterpart. The western jet decelerated quite fast, requiring
a local dense environment, but if the ISM is dense everywhere, the jet will be unable
to travel that far from the central BH. As a result, we consider a model that the ISM
density varies as the distance changes. For simplicity, we test the ideal case that the
jet travelled first through a ``cavity" with a constant velocity and then through a
dense region and was decelerated there. A new parameter $r$, the outer radius of the
cavity, is introduced and the ISM number density is set to be a constant $n$ outside
this region and zero inside. The fittings improved a lot but not well constrained
because of the limited number of the data points. A combination of Lightcurve fitting
is required to help the determination.

\subsection{Radiation Model}
In the standard GRB scenario, the afterglow emission is produced by
the synchrotron radiation or inverse Compton emission of the
accelerated electrons in the shock front of the jets (Wang et al.
2003 and references there). Wang et al.(2003) found that the reverse
shock emission, originating from the electrons of the jet when a
shock moves back through the ejecta, decay rather fast and describe
the data of the eastern jet quite well. We thus take this model in
our work as well.

Assuming the distribution of the electrons obeys a power-law form,
$n{\gamma_{\textrm{\tiny e}}}d\gamma_{\textrm{\tiny
e}}=K\gamma_{\textrm{\tiny e}}^{-p}d\gamma_{\textrm{\tiny e}}$, for
$\gamma_{\textrm{\tiny m}}<\gamma_{e}<\gamma_{\textrm{\tiny M}}$,
the volume emissivity at frequency $\nu'$ in the comoving frame is
given by
\begin{equation}
j_{\nu'}=\frac{\sqrt{3}q^{3}}{2m_{\textrm{\tiny
e}}c^{2}}(\frac{4{\pi}m_{\textrm{\tiny
e}}c\nu'}{3q})^{\frac{(1-p)}{2}}B_{\pm}^{\frac{(p+1)}{2}}KF_{1}(\nu,\nu'_{\textrm{\tiny
m}},\nu'_{\textrm{\tiny M}}),
\end{equation}
where $F_{1}(\nu,\nu'_{\textrm{\tiny m}},\nu'_{\textrm{\tiny
M}})=\int_{\nu'/\nu'_{\textrm{\tiny M}}}^{\nu'/\nu'_{\textrm{\tiny
m}}}F(x)x^{(p-3)/2}dx$, with $F(x)=x\int_{0}^{+\infty}K_{5/3}(t)$
and $K_{5/3}(t)$ is the Bessel function. The physical quantities in
these equations include $q$ and $m_{\textrm{\tiny e}}$, the charge
and mass of the electron, $B_{\perp}$, the magnetic field strength
perpendicular to the electron velocity, and $\nu'_{\textrm{\tiny
m}}$ and $\nu'_{\textrm{\tiny M}}$, the characteristic frequencies
for electrons with $\gamma_{\textrm{\tiny m}}$ and
$\gamma_{\textrm{\tiny M}}$.

Assuming the reverse shock heats the ejecta at time $t_{0}$ at the
radius $R_{0}$, the physical quantities in the adiabatically
expanding ejecta with radius $R$ will evolve as
$\gamma_{\textrm{\tiny m}}=\gamma_{\textrm{\tiny m}}(t_{0})R_{0}/R,
\gamma_{\textrm{\tiny m}}=\gamma_{\textrm{\tiny m}}(t_{0})R_{0}/R$
and $K=K(t_{0})(R/R_{0})^{-(2+p)},
B_{\perp}=B_{\perp}(t_{0})(R/R_{0})^{-2}$, where the initial values
of these quantities are free parameters to be fitted in the
calculation.

With these assumptions, we can then calculate the predicted flux
evolution of the jets. The comoving frequency $\nu'$ relates to our
observer frequency $\nu$ by $\nu=D\nu'$, where $D$ is the Doppler
factor and we have $D_{\textrm{\tiny
a}}=1/\Gamma(1-\beta\cos\theta)$ and $D_{\textrm{\tiny
r}}=1/\Gamma(1+\beta\cos\theta)$ for the approaching and receding
jets respectively. Considering the geometry of the emission region,
the observed X-ray flux in 0.3-8 keV band could be estimated by
\begin{equation}
F(\textrm{0.3-8
keV})=\int_{\nu_{1}}^{\nu_{2}}[\frac{\theta_{j}^{2}}{4}(\frac{R}{d}){\Delta}RD^{3}j_{\nu'}]d\nu,
\end{equation}
where ${\Delta}R$ is the width of the shock region and is assumed to
be ${\Delta}R=R/10$ in the calculation.

To reduce the number of free parameters, we set
$\gamma_{\textrm{\tiny m}}=100$ in our calculation because the
results are quite insensitive to this value. We choose the time that
the reverse shock takes place according to our kinematic model in
section 3.1. Then we fit the data to find out the initial values of
$K$ and $B_{\perp}$.

Next step, we combine the kinematic and radiation fitting together.
We know that the energy and the number density of the gas in the
pre-shock and post-shock regions are connected by the jump
conditions $n'=\zeta(\Gamma)n$ and $e'=\eta(\Gamma)nm_{\textrm{\tiny
p}}c^{2}$, where $\zeta(\Gamma)$ and $\eta(\Gamma)$ are coefficients
related to the jet velocity. Therefore if we assume the shocked
electrons and the magnetic field acquire constant fractions
($\epsilon_{\textrm{\tiny e}}$ and $\epsilon_{\textrm{\tiny B}}$) of
the total shock energy, we have $\gamma_{\textrm{\tiny
m}}=\epsilon_{\textrm{\tiny e}}(p-2){m_{\textrm{\tiny
p}}}(\Gamma-1)/(p-1){m_{\textrm{\tiny e}}}$,
$K=(p-1)n'\gamma_{\textrm{\tiny m}}^{p-1}$, and
$B_{\perp}=\sqrt{8\pi\epsilon_{\textrm{\tiny B}}e'}$.

If we further assume that the $\epsilon_{\textrm{\tiny e}}$ of the
eastern and the western jets is the same, we may infer that
$K\propto{e'}\propto{n}$ for the two jets. As a result, we search
for the combination of parameters that could satisfy the kinematic
and radiation fitting, as well as the relationship $K_{\textrm{\tiny
e}}/K_{\textrm{\tiny w}}{\sim}n_{\textrm{\tiny e}}/n_{\textrm{\tiny
w}}$.

A set of parameters has finally been found (Please refer to the \textit{left} panel in
Fig.2). The boundary of the cavity lies at $r\sim$14 arcsec to the east and $\sim$18
arcsec to the west of the central source. The corresponding number density of the ISM
outside this boundary is $\sim$0.00675 cm$^{-3}$ and $\sim$0.21 cm$^{-3}$,
respectively. These values are both lower than the canonical ISM value of $\sim$1
cm$^{-3}$, although the value in the western region is much higher than in the eastern
region. The electron energy fraction relationship is satisfied as $K_{\textrm{\tiny
e}}/K_{\textrm{\tiny w}}{\sim}n_{\textrm{\tiny e}}/n_{\textrm{\tiny w}}\sim0.03$. But
the other relation concerning the magnetic field strength could not be satisfied
simultaneously by these parameters. Although the cavity radius and the number density
are allowed to vary significantly, the best fitted magnetic field strength remains
quite stable($\sim$0.4-0.6 mG). One possible interpretation for this is that the
equipartition parameter varies as the physical conditions of the jet varies; an
alternative explanation may involve the {\it in situ} generation (or amplification) of
the magnetic field.

\addtocounter{figure}{1}
\begin{figure}
\begin{minipage}[b]{.5\textwidth}
    \centering
    \includegraphics[scale=0.4]{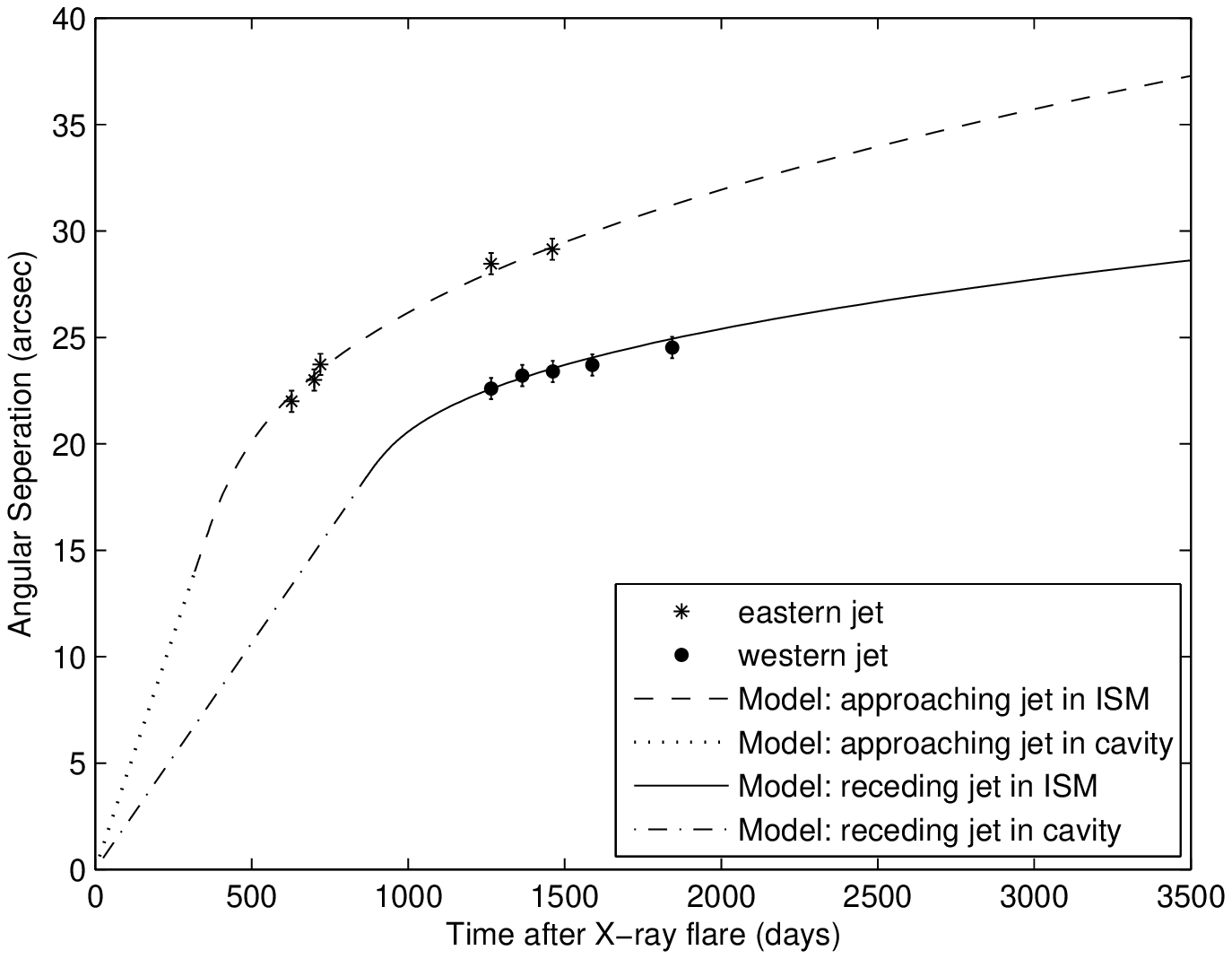}
  \end{minipage}
   \begin{minipage}[b]{.5\textwidth}
    \centering
    \includegraphics[scale=0.4]{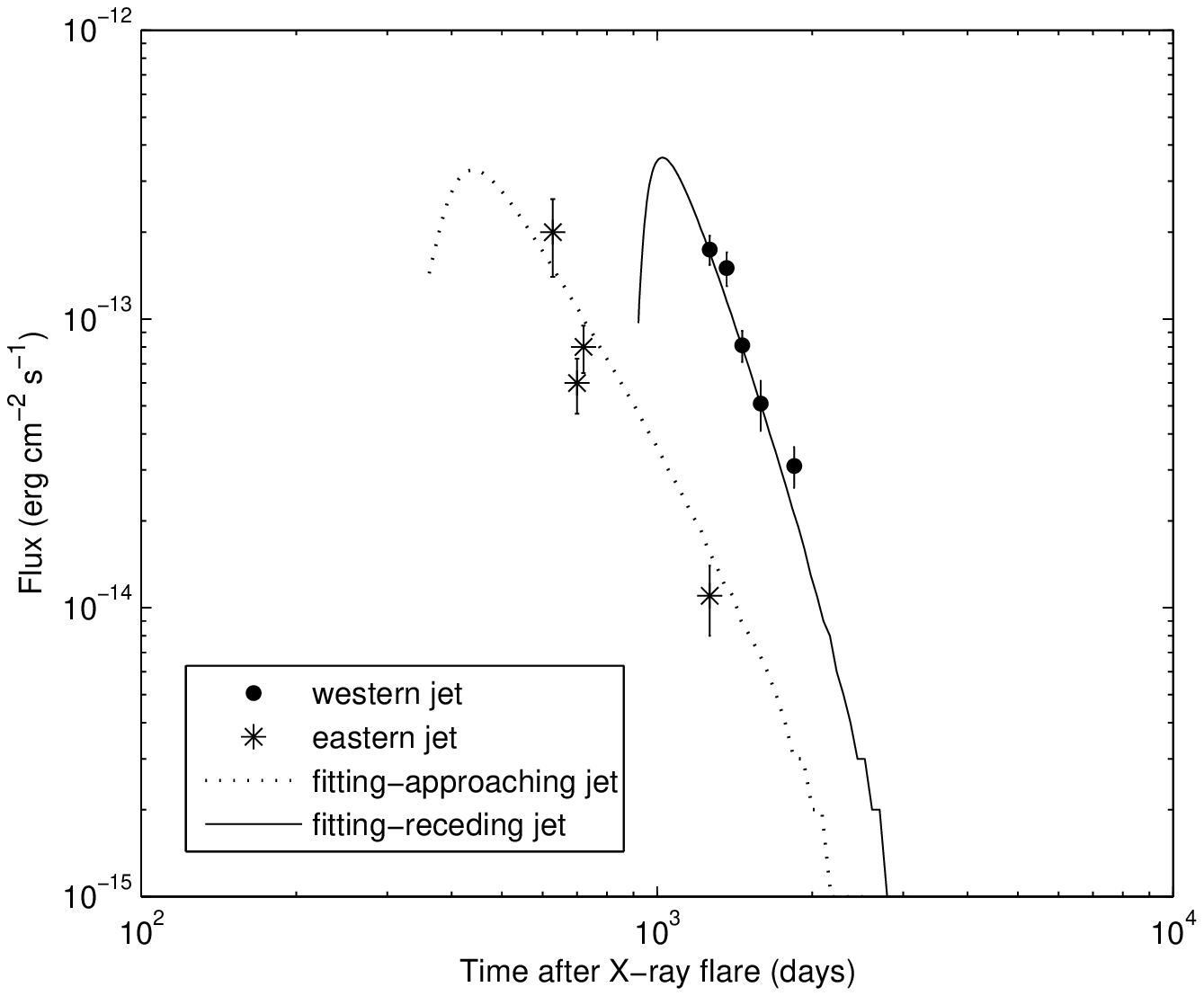}
  \end{minipage}
  \begin{minipage}[t]{.99\textwidth}
  \caption{{\it Left}: Proper motion fitting with asymmetric
cavity. Parameters: $\theta$=68\textordmasculine, $r_{\textrm{\tiny
e}}$=14 arcsec, $r_{\textrm{\tiny w}}$=18 arcsec, $n_{\textrm{\tiny
e}}$=0.00675 cm$^{-3}$, $n_{\textrm{\tiny w}}$=0.21 cm$^{-3}$. {\it
Right}: Reverse shock emission fitting to the X-ray light curve of
the jets. $K_{\textrm{\tiny e}}$=0.09 cm$^{-3}$, $K_{\textrm{\tiny
w}}$=0.3 cm$^{-3}$, $B_{\textrm{\tiny e}}$=0.5 mG, $B_{\textrm{\tiny
w}}$=0.4 mG.}
 \end{minipage}
\end{figure}

\section{Conclusion and Discussions}
External shock model shows that a large scale cavity exists outside XTE J1550-564. This
model has also been applied to another X-ray transient H 1743-322. Chandra X-ray and
ATCA radio observations of this source from 2003 November to 2004 June revealed the
presence of large-scale ($\sim$0.3 pc) jets with velocity $v/c\sim0.8$ (Rupen et al.
2004; Corbel et al. 2005). Deceleration is also confirmed in this system. The external
shock model describes the data of this source consistently. A cavity of size $\sim$0.12
pc is likely to exist, but not very clear in this case. Even if there is no vacuum
cavity, the ISM density is found to be extremely low($\sim3\times10^{-4}$ cm$^{-4}$),
compared to the canonical Galactic value.

These studies led us to the suggestion that in microquasars the
interactions between the ejecta and the environmental gas play major
roles in the jet evolution and the low density of the environment is
a necessary requirement for the jet to develop to a long distance.

When putting together all the analyses of microquasar jets, we found that microquasar
jets can be classified into roughly three groups: small scale moving jets, large scale
moving jets and large scale jet relics. For the first type, the ``small jets", only
radio emissions are detected. The jets are always relatively close to the central
source and dissipate very quickly, including GRS 1915+105 (Rodr\'{\i}guez \& Mirabel
1999; Miller-Jones et al. 2007), GRO J1655-40 (Hjellming \& Rupen 1995), and Cyg X-3
(Marti et al. 2001). The typical spatial scale is 0$\sim$0.05 pc and the time scale is
several tenths of days. No obvious deceleration is observed before the jets become too
faint. For the second type, the ``large jets", both X-ray and radio detections are
obtained, at a place far from the central source several years after the outburst.
Examples are XTE J1550-564, H1743-322, and GX 339-4 (Gallo et al.2004). The typical jet
travelling distance for this type is 0.2$\sim$0.5 pc from the central engine and
deceleration is clearly observed. The last type, the ``large relics", is a kind of
diffuse structures observed in radio, optical and X-ray band, often ring or nebula
shaped that are not moving at all. In this class, some well studied sources, Cygnus X-1
(Gallo et al.2005), SS433 (Dubner el al.1998), Circinus X-1 (Stewart et al. 1993) and
GRS1758-258 (Rodr\'{\i}guez et al. 1992) are included. The typical scale for this kind
is 1$\sim$30 pc, an order of magnitude larger than the second type. The estimated
lifetime often exceeds one million years, indicating that they are related to previous
outbursts.

From these properties, it is reasonable to further suggest a consistent picture
involving all the sources together. We make a conjecture that large scale cavities,
exist in all microquasar systems. The ``small jets" observed right after the ejection
are just travelling through these cavities. Since there are few or none interactions
between the jets and the surrounding gas in this region, the jets travel without
obvious deceleration. The emission mechanism is synchrotron radiation by particles
accelerated in the initial outburst. The emissions of jets decay very quickly and are
not detectable after several tenths of days. In some cases (e.g. XTE J1550-564), the
cavity has a dense (compared to the cavity) boundary at some radius and the
interactions between the jets and the boundary gas heat the particles again and thus
make the jets detectable again. Those are the ``large jets". The emission mechanism
then is synchrotron radiation by the re-heated particles in the external shocks. Then,
after these interactions, the jets lost most of their kinetic energy into the ISM
gradually, causing the latter to expand to large scale structures, the ``large relics",
in a comparatively long time (several millions of years).

The creation of the cavities is not clear at this stage. Possible mechanism may involve
supernova explosion, companion star winds or disk winds. Since some of the sources most
likely never had supernovae before and the winds from the companion stars are not
strong enough, the accretion disk winds may be the most plausible possibility. However,
these assumptions all require further observations to justify.

Microquasars are powerful probes of both the central engine and
their surrounding environment. More studies of the jets behaviors
may give us information on the ISM gas properties, as well as the
ejecta components. It will provide insights of the jet formation
process and offer another approach into black hole physics and
accretion flow dynamics.


\begin{theacknowledgments}
 We thank Dr. Yuan Liu, Shichao Tang and Weike Xiao for
useful discussions and Xiangyu Wang for providing the model codes. SNZ is grateful to
Prof. Sandip Chakrabarti for his great effort in organizing this conference, and to the
great hospitality of S.N. Bose National Centre for Basic Sciences, Kolkata, India. SNZ
acknowledges partial funding support by the Yangtze Endowment from the Ministry of
Education at Tsinghua University, Directional Research Project of the Chinese Academy
of Sciences under project No. KJCX2-YW-T03 and by the National Natural Science
Foundation of China under project no. 10521001, 10733010 and 10725313.
\end{theacknowledgments}



\bibliographystyle{aipproc}   

\bibliography{sample}

\begin{thebibliography}{20}
\bibitem[Corbel et al.(2001)]{cor01} Corbel S., Kaaret P., Jain R.K.,et al. 2001, ApJ, 554, 43,
\bibitem[Corbel et al.(2002)]{cor02} Corbel, S., Fender, P. R., et al. 2002, Science, 298, 196
\bibitem[Corbel et al.(2005)]{cor05} Corbel, S., Kaaret, P., et al. 2005, ApJ, 632, 504
\bibitem[Corbel et al.(2006)]{cor06} Corbel, S, Tomsick, J. A, \& Kaaret, P, 2006, ApJ, 636, 971
 \bibitem[Dickey and Lockman(1990)]{dic90} Dickey, J. M., \& Lockman, F. J. 1990, ARA\&A, 28, 215
 \bibitem[Dubner et al.(1998)]{dub98} Dubner, G. M., Holdaway, M., Goss, W. M., \& Mirabel, I. F., 1998, ApJ 116,1842
\bibitem[Fender et al.(2004)]{fen04} Fender, R.P., Belloni, T.£Í, \& Gallo, E., 2004, MNRAS, 355,1105
\bibitem[Fender et al.(2004)]{fend04} Fender, R.P., Gallo, E.,\& Jonker, P., 2004, Nucl.Phys. B, 132,346
 \bibitem[Freeman et al. (2002)]{fre02} Freeman, P. E., Kashyap, V., Rosner, R., \& Lamb, D. Q. 2002, ApJS, 138, 185
 \bibitem[Gallo et al.(2004)]{gal04} Gallo, E., Corbel, S., Fender, R. P., et al., 2004, MNRAS, 347, L52
 \bibitem[Gallo et al.(2005)]{gal05} Gallo, E., Fender, R. P., Kaiser, C., et al. 2005, arXiv:astro-ph/0508228v1
 \bibitem[Hannikainen et al.(2001)]{han01} Hannikainen, D., Campbell-Wilson, D., Hunstead, R., et al. 2001, ApSS Supp., 276, 45
\bibitem[Heinz(2002)]{hei02} Heinz, S., 2002, AA, 388, L40
 \bibitem[Hjellming(1995)]{hje95} Hjellming, R.M., \& Rupen, M. P., 1995, Nature, Vol.375, 464
 \bibitem[Huang et al.(1999)]{hua99} Huang, Y. F., Dai, Z.G., \& Lu, T. 1999, MNRAS, 309, 513
 \bibitem[Karret et al.(2003)]{kar03} Karret, P, Corbel, S, \& Tomsick, J.A, 2003, ApJ, 582, 945
 \bibitem[Marti et al.(2001)]{mar01} Marti, J., Paredes, J., M., \& Peracaula, M., 2001, A\&A 375,476
 \bibitem[Miller-Jones et al.(2007)]{mil07} Miller-Jones, J. C. A., Rupen, M. P., Fender, R. P., et al., 2007, MNRAS, 375,1087
 \bibitem[Mirabel et al.(1993)]{mir93} Mirabel, I. F., Rodr\'{\i}guez, L. F., Cordier B., et al. 1993, A\&A, suppl.Ser., 97, 193
\bibitem[Mirabel and Rodr\'{\i}guez(1994)]{mir94} Mirabel, I. F., \& Rodr\'{\i}guez, L. F. 1994, Nature, 371, 46
\bibitem[Mirabel and Radr\'{\i}guez(1999)]{mir99} Mirabel, I. F., \& Rodr\'{\i}guez, L. F. 1999, ARA\&A, 37, 409
\bibitem[Mirabel and Radr\'{\i}guez(2003)]{mir03} Mirabel, I. F., \& Rodr\'{\i}guez, L. F. 2003, Science, Vol300, 1119
 \bibitem[Orosz et al.(2002)]{oro02} Orosz, J. A., Groot, P. J., van der Klis, M., et al., 2002, ApJ, 568, 845
 \bibitem[Rodr\'{\i}guez et al.(1992)]{rod92} Rodriguez, L. F., \& Mirabel, I. F., \& Marti, J. 1992, ApJ, 401,L15
 \bibitem[Rupen et al.(2004)]{rup04} ------- 2004, BAAS, 204, 5.16
 \bibitem[Smith(1998)]{smi98} Smith, D. A., 1998, Int. Astron. Union Circ. No. 7008
\bibitem[Sobczak et al.(2000)]{sob00} Sobczak, G. J, McClintock, J. E, et al., 2000, ApJ, 544,993
\bibitem[Stewart et al.(1993)]{ste93} Stewart, R. T., Caswell, J. L., Haynes, R. F., \& Nelson, G. J., 1993, MNRAS, 261, 593
\bibitem[Stirling et al.(2001)]{sti01} Stirling, A. M., Spencer, R. E., de la Force, C. J., et al. 2001, MNRAS, 327, 1273
\bibitem[Sturner and Shrader(2005)]{stu05} Sturner S. J., \& Shrader, C. R. 2005, ApJ, 625,923
\bibitem[Tingay et al.(1995)]{tin95} Tingay, S. J., Jauncey, D. L., Prestonet, R. A., et al. 1995, Nature, 374, 141
\bibitem[Tomsick et al.(2003)]{tom03} Tomsick, J. A., Corbel, S., \& Fender, R. 2003, ApJ, 582,933
\bibitem[Wang et al.(2003)]{wan03} Wang, X. Y., Dai, Z. G., \& Lu, T. 2003, ApJ, 592,347
\end{thebibliography}

\IfFileExists{\jobname.bbl}{}
 {\typeout{}
  \typeout{******************************************}
  \typeout{** Please run "bibtex \jobname" to optain}
  \typeout{** the bibliography and then re-run LaTeX}
  \typeout{** twice to fix the references!}
  \typeout{******************************************}
  \typeout{}
 }


\end{document}